\documentclass[letter,longauth,traditabstract]{aa} 
\usepackage{graphicx}
\usepackage{natbib}
\bibpunct{(}{)}{;}{a}{}{,} 
\usepackage{txfonts}
\begin{document}
   \title{Deep {\it Herschel} view of obscured star formation in the Bullet cluster\thanks{{\it Herschel} is an ESA space observatory with science
       instruments provided by European-led Principal Investigator
       consortia and with important participation from
       NASA.  Data presented in this paper were analyzed using
       ``The Herschel Interactive Processing Environment (HIPE),'' a
       joint development by the Herschel Science Ground Segment
       Consortium, consisting of ESA, the NASA Herschel Science
       Center, and the HIFI, PACS and SPIRE consortia.}}

  \author{
          T.~D.~Rawle\inst{\ref{inst1}}
          \and
         S.~M.~Chung\inst{\ref{inst19}}
         \and
          D.~Fadda\inst{\ref{inst9}}
          \and
          M.~Rex\inst{\ref{inst1}}
          \and
          E.~Egami\inst{\ref{inst1}}
          \and
          P.~G.~P\'{e}rez-Gonz\'{a}lez\inst{\ref{inst14},\ref{inst1}}
          \and
          B.~Altieri\inst{\ref{inst2}}
          \and
          A.~W.~Blain\inst{\ref{inst3}}
          \and
          C.~R.~Bridge\inst{\ref{inst3}}
          \and
          A.~K.~Fiedler\inst{\ref{inst1}}
          \and
         A.~H.~Gonzalez\inst{\ref{inst19}}
         \and
          M.~J.~Pereira\inst{\ref{inst1}}
          \and
          J.~Richard\inst{\ref{inst15}}
          \and
          I.~Smail\inst{\ref{inst15}}
          \and
          I.~Valtchanov\inst{\ref{inst2}}
          \and
          M.~Zemcov\inst{\ref{inst3},\ref{inst4}}
          \and
          P.~N.~Appleton\inst{\ref{inst9}}
	\and
          J.~J.~Bock\inst{\ref{inst3},\ref{inst4}}
          \and
          F.~Boone\inst{\ref{inst5},\ref{inst7}}
          \and
          B.~Clement\inst{\ref{inst6}}
          \and
          F.~Combes\inst{\ref{inst7}}
          \and
          C.~D.~Dowell\inst{\ref{inst3},\ref{inst4}}
          \and
          M.~Dessauges-Zavadsky\inst{\ref{inst8}} 
          \and
          O.~Ilbert\inst{\ref{inst6}}
          \and
          R.~J.~Ivison\inst{\ref{inst10},\ref{inst11}}
          \and
          M.~Jauzac\inst{\ref{inst6}}
          \and
          J.-P.~Kneib\inst{\ref{inst6}}
          \and
          D.~Lutz\inst{\ref{inst12}}
          \and
          R.~Pell\'{o}\inst{\ref{inst5}}
          \and
          G.~H.~Rieke\inst{\ref{inst1}}
          \and
          G.~Rodighiero\inst{\ref{inst16}}
          \and
          D.~Schaerer\inst{\ref{inst8},\ref{inst5}}
          \and
          G.~P.~Smith\inst{\ref{inst17}}
          \and
          G.~L.~Walth\inst{\ref{inst1}}
          \and
          P.~van~der~Werf\inst{\ref{inst18}}
          \and
          M.~W.~Werner\inst{\ref{inst4}}
          }

   \institute{Steward Observatory, University of Arizona,
              933 N. Cherry Ave, Tucson, AZ 85721, USA;
              \email{eegami@as.arizona.edu}\label{inst1}
          \and
         Department of Astronomy, University of Florida, Gainesville,
         FL 32611-2055, USA\label{inst19}
         \and
         NASA Herschel Science Center, California Institute of
         Technology, MS 100-22, Pasadena, CA 91125, USA\label{inst9}
         \and
         Departamento de Astrof\'{\i}sica, Facultad de
         CC. F\'{\i}sicas, Universidad Complutense de Madrid, E-28040
         Madrid, Spain\label{inst14}
         \and
         Herschel Science Centre, ESAC, ESA, PO Box 78, Villanueva de
         la Ca\~nada, 28691 Madrid, Spain\label{inst2}
         \and
         California Institute of Technology, Pasadena, CA 91125,
         USA\label{inst3}
         \and
         Institute for Computational Cosmology, Department of Physics,
         Durham University, South Road, Durham DH1 3LE, UK\label{inst15}
         \and
         Jet Propulsion Laboratory, Pasadena, CA 91109, USA\label{inst4}
         \and
         Laboratoire d'Astrophysique de Toulouse-Tarbes,
         Universit\'{e} de Toulouse, CNRS, 14 Av. Edouard Belin, 31400
         Toulouse, France\label{inst5}
         \and
         Laboratoire d'Astrophysique de Marseille, CNRS -
         Universit\'{e} Aix-Marseille, 38 rue Fr\'{e}d\'{e}ric
         Joliot-Curie, 13388 Marseille Cedex 13, France\label{inst6}
         \and
         Observatoire de Paris, LERMA, 61 Av. de l'Observatoire, 75014
         Paris, France\label{inst7}
         \and
         Geneva Observatory, University of Geneva, 51, Ch. des
         Maillettes, CH-1290 Versoix, Switzerland\label{inst8}
         \and
         UK Astronomy Technology Centre, Science and Technology
         Facilities Council, Royal Observatory, Blackford Hill,
         Edinburgh EH9 3HJ, UK\label{inst10}
         \and
         Institute for Astronomy, University of Edinburgh, Blackford
         Hill, Edinburgh EH9 3HJ, UK\label{inst11}
         \and
         Max-Planck-Institut f\"{u}r extraterrestrische Physik,
         Postfach 1312, 85741 Garching, Germany\label{inst12}
         \and
         Department of Astronomy, University of Padova,
         Vicolo dell'Osservatorio 3, I-35122 Padova, Italy\label{inst16}
         \and
         School of Physics and Astronomy, University of Birmingham,
         Edgbaston, Birmingham, B15 2TT, UK\label{inst17}
         \and
         Sterrewacht Leiden, Leiden University, PO Box 9513, 2300 RA
         Leiden, the Netherlands\label{inst18}
         }

  \date{Received April 1, 2010; accepted May 11, 2010}

 
  \abstract
  {{We use deep, five band (100--500 $\mu$m) data from the {\it Herschel} Lensing Survey (HLS) to fully constrain the obscured star formation rate, SFR$_{\rm FIR}$, of galaxies in the Bullet cluster ($z$ = 0.296), and a smaller background system ($z$ = 0.35) in the same field. {\it Herschel} detects 23 Bullet cluster members with a total SFR$_{\rm FIR}$ = 144 $\pm$ 14 M$_{\sun}$ yr$^{-1}$. On average, the background system contains brighter {far-infrared} (FIR) galaxies, with $\sim$50\% higher SFR$_{\rm FIR}$ (21 galaxies; 207 $\pm$ 9 M$_{\sun}$ yr$^{-1}$). SFRs extrapolated from 24 $\mu$m flux via recent templates (SFR$_{24{\mu}m}$) agree well with SFR$_{\rm FIR}$ for $\sim$60\% of the cluster galaxies. In the remaining $\sim$40\%, SFR$_{24{\mu}m}$ underestimates SFR$_{\rm FIR}$ due to a significant excess in observed $S_{100}/S_{24}$ (rest frame $S_{75}/S_{18}$) compared to templates of the same FIR luminosity.}}


   \maketitle

\section{Introduction}

In the last decade many studies have attempted to quantify the star formation rate (SFR) within cluster galaxies. Ultraviolet and optical observations have successfully identified trends between unobscured star formation and local environment, suggesting that star formation in cluster core galaxies is generally more quenched \citep[e.g.][]{kod04-1103,por07-1409}. However, star formation can be obscured by dust, which re-emits stellar light in the far-infrared (FIR), peaking at a rest frame $\lambda_0$ $\sim$ 100 $\mu$m. Mid-infrared surveys \citep[e.g.][]{met05-425,gea06-661,fad08-9} have explored obscured star formation by estimating total FIR luminosity from  template spectra. These templates are often based on small numbers of well constrained local galaxies, e.g. \citet{rie09-556}.

The PACS \citep{pog10} and SPIRE \citep{gri10} instruments, onboard the ESA {\it Herschel} Space Observatory \citep{pil10}, enable unprecedented multi-band coverage of the FIR. The {\it Herschel} Lensing Survey ({HLS}; PI: E Egami) consists of 5-band observations (100--500 $\mu$m) of 40 nearby clusters ($z$ $\sim$ 0.2--0.4). Nominally devised to exploit the gravitational lensing effect of massive clusters to observe high redshift galaxies (see \citealt{ega10} for details on survey design), a useful by-product is deep FIR observations of the clusters themselves. At these redshifts, {\it Herschel} photometry spans the peak of the dust component, allowing an accurate constraint of far infrared luminosity, {$L_{\rm FIR}$}, and hence obscured SFR.

During the {\it Herschel} Science Demonstration Phase, {HLS} observed the Bullet cluster (1E0657--56; $z$ = 0.296). The reason for this choice was two-fold. First, previous studies report bright {submillimeter} galaxies in the background \citep[e.g. ][]{rex09-348}, {with HLS analysis presented in \citet{rex10}}. Second, the Bullet cluster is a recent collision of two clusters \citep{mar02-27}, offering a unique laboratory for the study of star formation within a dynamic environment. The sub-cluster has conveniently fallen through the main cluster perpendicular to the line of sight ($< 8^\circ$ from the sky plane; \citealt{mar04-819}). Analysis of X-ray emission shows that a supersonic bow shock precedes the hot gas, while the weak lensing mass profile indicates that this X-ray bright component lags behind the sub-cluster galaxies due to ram pressure \citep{mar02-27,bar02-816}. A recent mid-infrared study by \citet{chu09-963} concluded that ram pressure from the merger event had no significant impact on the star formation rates of nearby galaxies. We can re-evaluate these previous studies by using {\it Herschel} data to constrain {$L_{\rm FIR}$} directly. In this letter, we present an exploration of obscured star formation in this cluster environment.

\begin{figure}
\centering
\includegraphics[height=80mm,angle=270,clip]{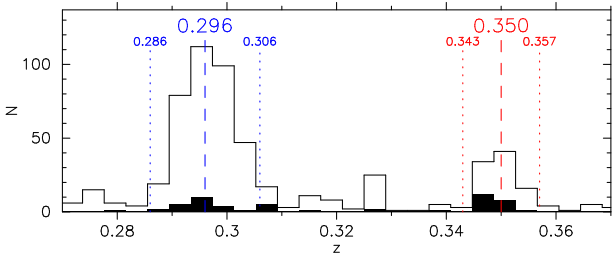}
\caption{Distribution of spectroscopic redshifts (0.27 $<$ $z$ $<$ 0.37) for galaxies within the Bullet cluster field (outline). {\it Herschel} detected galaxies are also shown (filled). In addition to the Bullet cluster ($z$ = 0.296), there is a background {system} at $z$ = 0.350. Dotted lines show our membership limits of 3000 km s$^{-1}$ and 2000 km s$^{-1}$ respectively.}
\label{fig:zhist}
\end{figure}

\section{Observations}

\subsection{Photometric data}

Five band {\it Herschel} imaging was obtained using two instruments: PACS (100,160 $\mu$m) covering approximately 8\arcmin$\times$8\arcmin~and SPIRE (250,350,500 $\mu$m) with a wider $\sim$17\arcmin$\times$17\arcmin~field. {We also use Magellan IMACS optical, \textit{Spitzer} IRAC and MIPS 24 $\mu$m maps with similar coverage to SPIRE, and high resolution \textit{HST} ACS images of the central 4\arcmin$\times$4\arcmin.} \citet{ega10} provides details of all the data, and presents {\it Herschel} FIR maps.

The deep SPIRE maps have detection limits well below the instrument confusion limits. To avoid compiling sourcelists from confused maps, {\it Herschel} fluxes are measured at all {\it Spitzer} MIPS 24 $\mu$m source positions. {For a typical galaxy SED at z $\sim$ 0.3, the 24 $\mu$m map is much deeper than SPIRE, so even with a relatively high S/N $>$ 10 cut (flux limit $\sim$100 $\mu$Jy), we can assume the inclusion of all sources contributing significant FIR flux.} The use of mid-infrared source positions has the added advantage of decreasing the significance of flux boosting, which has not been addressed in this study.

Photometric analysis followed the same procedure in all 5 {\it Herschel} bands. An average PSF, measured from the brightest unblended sources in the image, was simultaneously fit to all positions in the 24 $\mu$m catalogue (without re-centering) using {\sc Daophot Allstar}. At the longer SPIRE wavelengths, there is a higher probability of more than one 24 $\mu$m source falling within the FIR beam. In these instances, the objects are grouped together at the 24 $\mu$m S/N-weighted mean position, treated as a single source, and flagged (see following sub-section). For more details on the photometry technique see \citet{rex10}.

\subsection{Spectroscopy and sample selection}

The spectroscopic redshift catalogue combines observations from three campaigns: Magellan IMACS multi-slit (856 targets; \citealt{chu10s}, Chung et al. in prep), CTIO Hydra multi-fiber (202; Fadda et al. in prep) and VLT FORS multi-slit (14; J Richard, private communication). \citet{ega10} provides further details. The merged catalogue comprises 929 sources within the SPIRE field.

Figure \ref{fig:zhist} presents the distribution of spectroscopic redshifts for the range 0.27 $<$ $z$ $<$ 0.37. An important aspect of this study is confidence in the cluster membership of galaxies. The Bullet cluster distribution peaks at $z$ = 0.296, and we limit membership to $\pm$ 3000 km s$^{-1}$ (0.286 $<$ $z$ $<$ 0.306). In addition, this study also analyzes galaxies from a {system} at $z$ = 0.350 in the same field, limiting membership to $\pm$ 2000 km s$^{-1}$ (0.343 $<$ $z$ $<$ 0.357). {The systems have 362 and 95 known members respectively.}

The sample for this analysis consists of MIPS 24 $\mu$m sources with spectroscopically confirmed cluster redshifts. These two catalogues were merged by identifying the closest 24 $\mu$m source, within the RMS pointing error of MIPS (1.4\arcsec), to the spectroscopic position. For sample members grouped during the FIR photometry (previous sub-section), we examined the optical and IRAC colours of each group member, identifying the likely source of the mid- and far-IR flux. In cases where the sample member was not considered to be the source, or when the situation was unclear, the object was rejected from the sample.

In the final sample, there are 47 confirmed Bullet cluster members, and an additional 28 sources in the $z$ = 0.35 system. Of these, 23 and 21 galaxies respectively are detected in the {\it Herschel} bands, highlighted by the filled distribution in Fig. \ref{fig:zhist}. The background system has a much higher fraction of {\it Herschel} detections than the Bullet cluster {(75\% of 24 $\mu$m sources, compared to 50\%)}.

\begin{figure*}
\centering
\includegraphics[height=175mm,angle=270,clip]{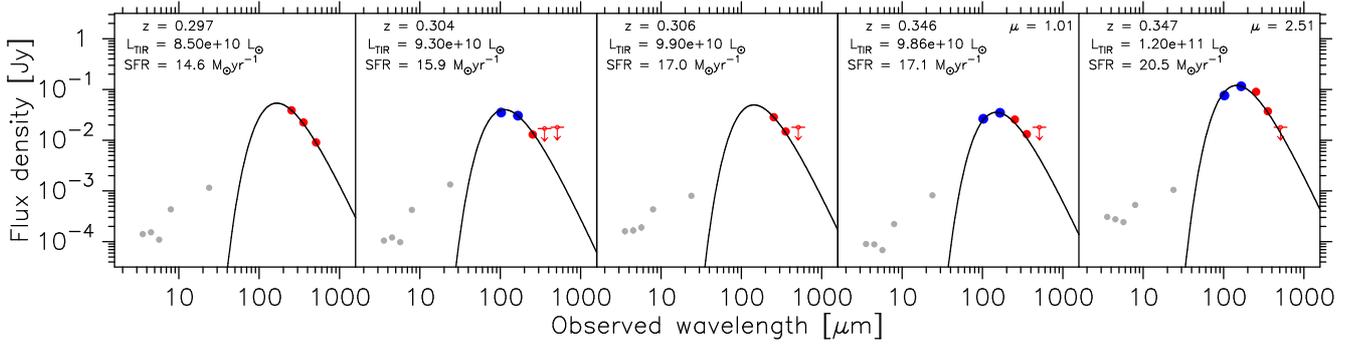}
\caption{Photometric data for five of the most FIR luminous galaxies in the sample. Blue = PACS; red = SPIRE; grey = IRAC/MIPS. Flux densities are as observed (i.e. not de-magnified). Redshift and, for background system galaxies, magnification factor, $\mu$, are displayed at the top of each panel. {$L_{\rm FIR}$} and SFR$_{\rm FIR}$ derived from the best fit blackbody (black line) are also shown and have been de-magnified where necessary.}
\label{fig:seds}
\end{figure*}

\section{Results and discussion}

\subsection{Far-infrared {(FIR)} spectral energy distributions}

For each source, the FIR spectral energy distribution (SED) is fit to all available {\it Herschel} data points, taking into account the upper limits for non-detections. The dust component is modeled by a modified, single-temperature, blackbody
\begin{equation}
S_{\nu} = N({\nu}/{\nu_0})^{\beta}B_{\nu}(T)
\end{equation}
where $S_{\nu}$ is flux density, $\beta$ is dust emissivity index (fixed at 1.5; {using $\beta$ = 2.0 would vary {$L_{\rm FIR}$} by $<$ 15\% on average}) and $B_{\nu}(T)$ is the Planck blackbody radiation function for a source at temperature $T$. The shape of this optically thin (rather than thick) blackbody imitates the inclusion of a secondary (warm) dust component. As we are concerned only with {$L_{\rm FIR}$} and SFR, the parameterization of the data is the most important aspect, and $T$ is used purely as a fit parameter. Galaxies within the PACS field have well constrained fits, and $T$ is allowed to float freely. For those without PACS data ($\sim$40\%), $T_{\rm 0}$ has been forced to a narrow range centered on the mean value from the constrained SEDs (30 $\pm$ 1 K). {Forcing $T_{\rm 0}$ to a similarly narrow range about values $\sim$1$\,\sigma$ from the constrained mean, varies {$L_{\rm FIR}$} by $<$ 25\%. Bias in {$L_{\rm FIR}$} due to model priors is comparable in scale to systematics from instrument calibration.}

{{$L_{\rm FIR}$} is integrated over (rest frame) $\lambda_0$ = 8--1000 $\mu$m} from which SFR$_{\rm FIR}$ is derived using the \citet{ken98-189} relation. As an illustration, Fig. \ref{fig:seds} displays the FIR SED fits for five of the most luminous galaxies in the sample. These simple fits may underestimate {$L_{\rm FIR}$} by up to a factor of 1.8 \citep{rex10}, as they lack a mid-infrared component. Future analysis will fully account for additional components. For the purposes of this study, a blackbody fit is sufficient. The luminosities of galaxies in the background system have been de-magnified using the Bullet cluster lensing model of Paraficz et al. (in prep). The remaining figures in this paper present de-magnified values.

\begin{figure}
\centering
\includegraphics[height=85mm,angle=270,clip]{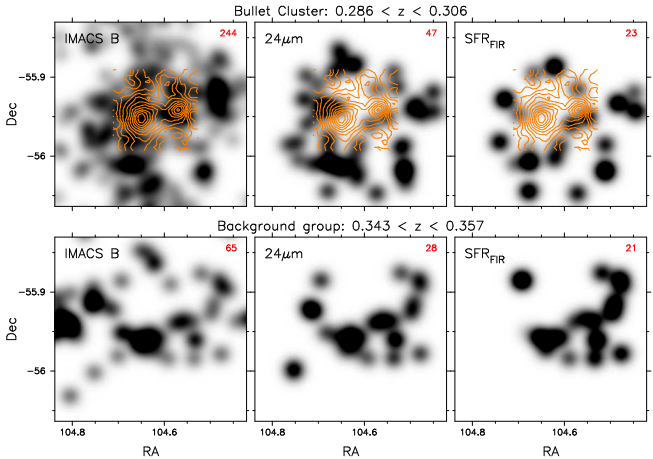}
\caption{Smoothed density maps for IMACS $B$-band flux (left panels), 24 $\mu$m flux (central) and SFR calculated from {\it Herschel} data (right). The sources are binned by confirmed system membership, with the number of contributing galaxies displayed in the upper-right of each panel: upper row for Bullet cluster (over-plot in orange by the weak lens mass map); lower row for $z$ = 0.35 {system}. All maps are Gaussian smoothed to the SPIRE 250 $\mu$m beam size (18\arcsec~FWHM).}
\label{fig:sfr}
\end{figure}

\subsection{Star formation rates within the systems}

The {system} at $z$ = 0.35 contains IR galaxies brighter than those in the Bullet cluster, with three galaxies meeting the LIRG criterion [log({$L_{\rm FIR}$}/$L_{\sun}$) $\ga$ 11.0] and an additional two within 1$\,\sigma$. A further 10 members have log({$L_{\rm FIR}$}/$L_{\sun}$) $>$ 10.5. In contrast, the Bullet cluster contains two LIRGs, and only six other galaxies brighter than log({$L_{\rm FIR}$}/$L_{\sun}$) = 10.5.

The total star formation rate of the 23 Bullet cluster galaxies is 144 $\pm$ 14 M$_{\sun}$ yr$^{-1}$. The 21 galaxies in the background {system} are, on average, $\sim$50\% more active, with a total SFR = 207 $\pm$ 9 M$_{\sun}$ yr$^{-1}$. Only five of these galaxies have been magnified by more than 20\%, and the minimum detected SFRs are similar in each system. Therefore, it is unlikely that the higher total SFR is due to a decreased lower limit caused by magnification. {The difference is likely to reflect the mass of the systems, although the lower SFR in the Bullet cluster may indicate that cluster--cluster mergers are not important for triggering FIR starbursts.}

Figure \ref{fig:sfr} displays the spatial distribution of the {\it Herschel}-derived SFR for the two systems. Flux densities in optical $B$-band and 24 $\mu$m are shown for comparison. {An initial examination suggests that the Bullet cluster exhibits a radial trend in SFR$_{\rm FIR}$ (lacking significant FIR detection towards the centre)}, reminiscent of that found in other {contemporary} studies \citep{bra10,per10}. The gradient in the Bullet cluster SFR is examined in detail in Chung et al. (in prep).

The IR and optical flux of the background {system} trace similar distributions, whereas in the Bullet cluster, the $B$-band flux is more centrally concentrated, away from the IR sources. This may indicate a different trend in dust retention for the two systems. While the 24 $\mu$m and {\it Herschel} SFR density maps generally trace the same distribution, there are significant outliers: bright 24 $\mu$m sources with relatively lower SFRs, and vice versa. In the following section, we compare the SFR estimated from 24 $\mu$m (through the \citealt{rie09-556} templates) to the SFR$_{\rm FIR}$.

\subsection{24$\mu$m as a {$L_{\rm FIR}$} predictor in nearby clusters}

The mid-infrared bands, e.g. MIPS 24 $\mu$m, are often used to estimate {far} infrared luminosity, {$L_{\rm FIR}$}, and hence obscured SFR, via template FIR SEDs such as \citet{rie09-556}. Those authors provide a simple formula (their equation 14) to convert 24 $\mu$m flux directly to SFR. The templates are based on local (U)LIRGs ($z$ $\la$ 0.1), and at high redshift, may not be valid. {Here, we test the template accuracy for cluster galaxies at $z$ = 0.3.}

\begin{figure}
\centering
\includegraphics[height=75mm,angle=270,clip]{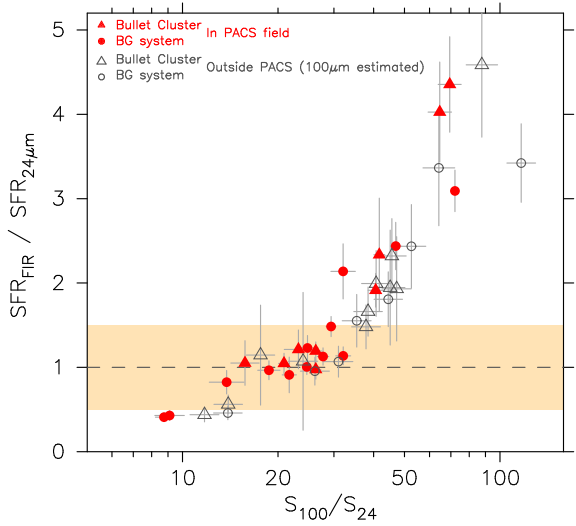}
\caption{{Ratio of SFR$_{\rm FIR}$ (from blackbody fit) to SFR$_{24{\mu}m}$ (via \citealt{rie09-556}) versus the flux ratio $S_{100}/S_{24}$. For galaxies outside the PACS field, 100 $\mu$m is predicted from the blackbody fit. Dashed line is equality and shaded region indicates 50\% difference in the SFRs. All galaxies with under-predicted SFR$_{24{\mu}m}$ have redder $S_{100}/S_{24}$.}}
\label{fig:24sfr}
\end{figure}

In Sect. 3.2 (Fig. \ref{fig:sfr}), we suggested that while 24 $\mu$m flux and SFR$_{\rm FIR}$ follow the same general distribution, they are not perfectly correlated. {A direct comparison of SFR$_{\rm FIR}$ to SFR$_{24{\mu}m}$ (Fig. \ref{fig:24sfr}; plotted against the dust-peak--mid-IR flux ratio) leads to the same conclusion. For $\sim$60\% of galaxies, the two SFRs agree well. However, there are several galaxies ($\sim$30\%) that have severely underestimated SFR$_{24{\mu}m}$, and these also display systematically redder $S_{100}/S_{24}$.} If SFR$_{\rm FIR}$ is underestimated by the simple blackbody fits (Sect. 3.1), the SFR$_{24{\mu}m}$ predictions are correspondingly worse.

\begin{figure}
\centering
\includegraphics[height=75mm,angle=270,clip]{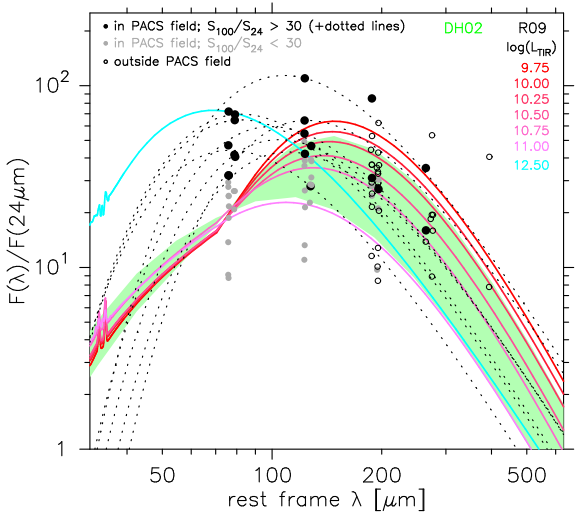}
\caption{Observed {\it Herschel} fluxes normalized by 24 $\mu$m for galaxies in both systems. {Symbols as described at top-left. \citet{rie09-556} average templates for 9.75 $\le$ log({$L_{\rm FIR}$}) $\le$ 11.0 (red--pink) plus one example high-{$L_{\rm FIR}$} template (cyan). Locus of low-activity templates from \citet{dal02-159} ($\alpha$ = 1.8--2.5) is shaded green. All templates are normalized at $\lambda_{\rm obs}$ = 24 $\mu$m ($\lambda_{\rm 0}$ = 18 $\mu$m). Templates in {$L_{\rm FIR}$} range of the observations under-predict $S_{100}/S_{24}$ for 40\% of sources. High {$L_{\rm FIR}$} templates do not match the shape at $\lambda$ $\ge$ 200 $\mu$m.}}
\label{fig:rieke}
\end{figure}

{Are the under-predicted SFR$_{24{\mu}m}$ caused by the redder $S_{100}/S_{24}$ colours?} Figure \ref{fig:rieke} examines the \citeauthor{rie09-556} templates more closely, comparing them to the {\it Herschel} fluxes. For templates spanning the {$L_{\rm FIR}$} range of the observations, the agreement is good for $\lambda_{\rm 0}$ $\ga$ 200 $\mu$m. However, at 100 $\mu$m there are 8 significant outliers; we define 100 $\mu$m excess galaxies as those with $S_{100}/S_{24}$ {(rest frame $S_{75}/S_{18}$)} $>$ 30 as the templates predict $S_{100}/S_{24}$ $\la$ 20. Only templates with very high luminosities, i.e. log({$L_{\rm FIR}$}/$L_{\sun}$) $\ge$ 12, match the observed $S_{100}/S_{24}$, but even the brightest sample galaxy has only log({$L_{\rm FIR}$}/$L_{\sun}$) $\sim$ 11.5, while most are log({$L_{\rm FIR}$}/$L_{\sun}$) $<$ 11. Although high {$L_{\rm FIR}$} templates have $S_{100}/S_{24}$ $\ga$ 30, their lower peak wavelength leads to an under-prediction at $\lambda$ $\ga$ 200 $\mu$m in at least three observed SEDs. {We also compare to the least active \citet{dal02-159} FIR templates ($\alpha$ = 1.8--2.5). The locus of these are substantially similar to the low luminosity \citeauthor{rie09-556} templates and thus also only under-predict $S_{100}/S_{24}$.} {We stress that, unlike $L_{\rm FIR}$ and SFR$_{\rm FIR}$, the presence of a 100 $\mu$m excess is independent of the blackbody fits and the systematic uncertainties therein.}

\begin{figure}
\centering
\includegraphics[width=90mm,clip]{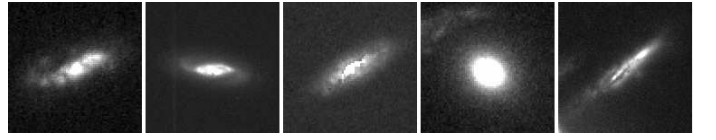}
\caption{HST ACS thumbnails of five 100 $\mu$m excess galaxies ($S_{100}/S_{24}$ $>$ 30; increasing from left) {do not suggest a single source population}.}
\label{fig:thumbs}
\end{figure}

Galaxies with a 100 $\mu$m excess account for $\sim$40\% of cluster members detected with PACS, and cover the entire range of {$L_{\rm FIR}$} sampled. Above a nominal luminosity limit of 10$^{10}$L$_{\sun}$, 55\% of Bullet cluster galaxies have the 100 $\mu$m excess. The fraction in the background {system} is lower at {36\%}. This may indicate a trend with environment, or could be due to the off-centre view of the latter system (i.e a potential radial trend). High resolution HST imaging covers five of the eight 100 $\mu$m excess galaxies (Fig. \ref{fig:thumbs}). Despite the small number, the galaxies span a broad range of types and morphologies. Further examples are required for a firm conclusion, but these suggest that the 100 $\mu$m excess is not due to a single population of galaxies.

The $S_{100}/S_{24}$ colours alone may have led to the conclusion that the 100 $\mu$m excess was due to galaxies with generally colder dust. However, fits to the combined {HLS} PACS+SPIRE photometry suggest that this is not the case. Rather, the excess may be due to an additional warm dust component or active galactic nuclei (AGN) which are not considered in the templates. Using a simple power law to parameterize flux in the range 24--100 $\mu$m, we estimate the AGN contribution to total bolometric luminosity via the $S_{60}/S_{25}$ indicator for ULIRGs \citep[fig. 36]{vei09-628}. None of the 100 $\mu$m excess galaxies have predicted AGN fractions $>$ 30\%. However, we may be under-predicting the contribution if the mid-IR SED steepens beyond 60 $\mu$m, or if the indicator breaks down for galaxies in this luminosity range.

{{\it Herschel} PACS observations of $z$ $\sim$ 0.2 LoCuSS clusters (without the advantage of complementary SPIRE data), display a similar fraction of 100 $\mu$m excess galaxies \citep{smi10,per10}. However, the high redshift field sample from the HLS Bullet cluster observations \citep{rex10} lacks a comparable excess at $\lambda_{\rm 0}$ $\approx$ 75 $\mu$m.} These results suggest that the effect could be either redshift dependent or cluster-specific. {HLS} is well placed for further analysis of the $S_{100}/S_{24}$ phenomenon, as the combined PACS+SPIRE data ensures that both the excess and entire FIR component can be constrained simultaneously.

\section{Conclusions}

Using deep {\it Herschel} observations (100--500 $\mu$m) to fully constrain the FIR component, we derive obscured SFRs for galaxies in the Bullet cluster ($z$ = 0.296), and a background system ($z$ = 0.35) in the same field. {{\it Herschel} detects 23 Bullet cluster members, with a total SFR$_{\rm FIR}$ = 144 $\pm$ 14 M$_{\sun}$ yr$^{-1}$, while the background system contains 21 detections but $\sim$50\% higher SFR (207 $\pm$ 9 M$_{\sun}$ yr$^{-1}$). The relative distributions of SFR$_{\rm FIR}$ and optical flux suggest a difference in dust retention between the two systems. For $\sim$60\% of galaxies, SFR$_{\rm FIR}$ agrees well with estimated SFRs from 24 $\mu$m flux via recent templates. However, the remaining galaxies display a significant excess at 100 $\mu$m ($\lambda_{\rm 0}$ $\approx$ 75 $\mu$m) compared to templates, which causes an under-prediction in SFR$_{24{\mu}m}$}. We note that such an excess is not found in the high redshift, field sample \citep{rex10}. Future studies will exploit the full range of 5-band {\it Herschel} cluster observations available in {HLS}, to form a more complete understanding of the environmental effect on obscured star formation rates, and explore the origin and dependencies of the 100 $\mu$m excess.

\begin{acknowledgements}
This work is based in part on observations made with {\it Herschel}, a
European Space Agency Cornerstone Mission with significant
participation by NASA.  Support for this work was provided by NASA
through an award issued by JPL/Caltech.
\end{acknowledgements}

\end{document}